\documentclass[prl,twocolumn,showpacs]{revtex4} 
\usepackage{epsfig}
\usepackage{amssymb}
\usepackage{color}
\newcommand{\figcap}[1]{\caption{#1}}
\newcommand{\bcen}{\begin{center}}
\newcommand{\ecen}{\end{center}}

\newcommand{\beq}{\begin{equation}}
\newcommand{\eeq}{\end{equation}}
\newcommand{\bea}{\begin{eqnarray}}
\newcommand{\eea}{\end{eqnarray}}
\newcommand{\non}{\nonumber}
\newcommand{\half}{\frac{1}{2}}
\newcommand{\bary}{\begin{array}}
\newcommand{\eary}{\end{array}}
\newcommand{\beps}{\mbox{\boldmath $ \epsilon $}}

\newcommand{\eqn}[1] {eqn.~(\ref{#1})}

\newcommand{\fig}[1]{fig.~\ref{#1}}
\newcommand{\Fig}[1]{Figure.~\ref{#1}}

\newcommand{\bn} { \mbox{\boldmath $n$}}
\newcommand{\bu} { \mbox{\boldmath $u$}}

\begin{document}
\title {Patterns, Forces and Metastable Pathways in Debonding of Elastic Films}
\author{Jayati Sarkar$^1$}
\author{Vijay Shenoy$^2$}
\email[]{shenoy@mrc.iisc.ernet.in}
\author{Ashutosh Sharma$^1$}
\email[] {ashutos@iitk.ac.in}
\affiliation{$^1$ Chemical Engineering Department, Indian Institute of Technology Kanpur,
UP 208 016, India \\ $^2$ Material Research Centre, Indian Institute of Science, Bangalore 560 012, India}
\begin{abstract}

The letter resolves several intriguing and fundamental aspects of
debonding at soft interfaces, including the formation and persistence
of regularly arranged nanocavities and bridges,
``adhesion-debonding hysteresis'', and vastly lower
adhesive strengths compared to the absence of pattern
formation.  The analysis shows the hysteresis to be
caused by an energy barrier that separates the metastable patterned
configurations during withdrawal, and the debonded state. The metastable
morphological pathways involving cavitation and peeling of contact
zones engender substantially lower debonding forces.

\end{abstract}
\pacs{82.35.Gh, 68.35.Ct, 68.55.-a, 46.50.+a}
\maketitle

The related phenomena of adhesion, debonding and interfacial cavitation
or cracking at soft elastic interfaces have been intensely studied both
in view of their technological applications and the many unresolved
scientific issues related to the pathways, morphology and forces of
debonding.  A rigid surface(contactor) initially in contact with a
soft elastic film, upon withdrawal, debonds by the formation of a
pattern of well defined spacing consisting of areas of intimate
contact and interfacial cavities\cite{1,2,3,4}. A linear stability
analysis\cite{5} showed spontaneous surface roughening of the film
when the contactor is initially brought in close proximity
($<$20nm). The wavelength ($\lambda$) of this surface pattern depends
only on the film thickness ($h$) ($\lambda \sim 3 h$), but is
independent of the strength and nature of the adhesive interactions as
well as the elastic properties of the film. The regions of adhesive
contact persist until the contactor is pulled to a much larger
distance than that at which the initial contact instability was
triggered upon approach. This difference between approach and
withdrawal behaviors may be referred to as ``adhesion-debonding or contact
hysteresis''. Another interesting aspect is the pull-off force
required for debonding can be several orders smaller than the force
calculated based on the assumption of flat surfaces. Clearly, such a
significant reduction cannot be explained merely by $\sim$50$\%$
reduction in the contact area observed at detachment\cite{3}. The
mechanism of ``adhesion-debonding hysteresis'', its associated morphologies
including patterned cavitation, pull-off force and distance are the
key unresolved issues in debonding at soft interfaces that are
addressed here.

\begin{figure}
\centerline{\epsfxsize=8.0truecm \epsfbox{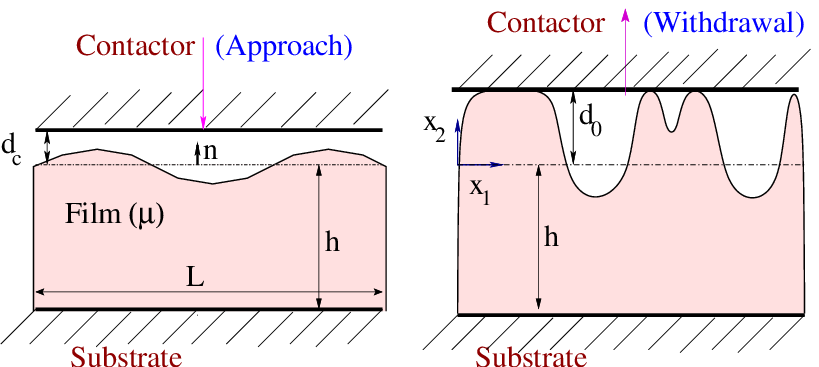}}
\figcap{Left: Schematic of small amplitude pattern formation on approach at critical distance $d_c$. Right: Column stretching and growth of cavities on withdrawal of the contactor.}
\label{figure1}
\end{figure}

\Fig{figure1} illustrates the film-contactor geometry. The total energy consists of the stabilizing stored elastic energy and the destabilizing attractive interaction between the contactor and the film.
\bea
\Pi =  \int_V W(\beps) dV + \int_S U(d_0-\bu \cdot \bn) dS, \label{TotalEnergy}
\eea
where $W$ is the strain energy density
defined as, $W(\beps)= \half \mu(\beps:\beps)$ with $\beps$ is the
strain tensor, $\bu$ is the displacement vector, $\mu$ is the shear
modulus of the film ($\lesssim$10MPa). The interaction potential consists of
an attractive van der Waals component along with a short range Born
repulsion, represented by $U(d_0-\bu \cdot \bn)=-A/12\pi(d_0-\bu \cdot
\bn)^2+B/(d_0-\bu\cdot\bn)^8$ where $A$ is the Hamaker constant (of
the order of $10^{-20}J$) and $B$ is the coefficient of Born
repulsion.  The coefficient $B$ is correlated to the adhesive energy
at contact $(\Delta G = U(d_e) = A/(16 \pi d_e^2) )$, where $d_e$ is the equilibrium
separation distance obtained from $U'(d_e) = 0$. This form of
interaction implies that the force required to pull off two rigid flat
surfaces is $F_{max}^{flat} = \Delta G/d_e$. As may be expected,
our detailed studies (to be published) have confirmed that the debonding is
controlled by the adhesive energy rather than the detailed functional
form of the potential. Based on the linear stability of
\eqn{TotalEnergy}, it was shown that the film surface becomes
spontaneously rough as the contactor approaches it to within a small
critical distance $d_c$ at which $h |U''|/\mu \ge 6.22$. The
lengthscale of the pattern consisting of cavities and bridges (regions
of contact) is about three times the film thickness regardless of the
interaction potential and the elastic properties; both of which are in
agreement with observations\cite{2,3,4}.

\begin{figure}
\centerline{\epsfxsize=8.0truecm \epsfbox{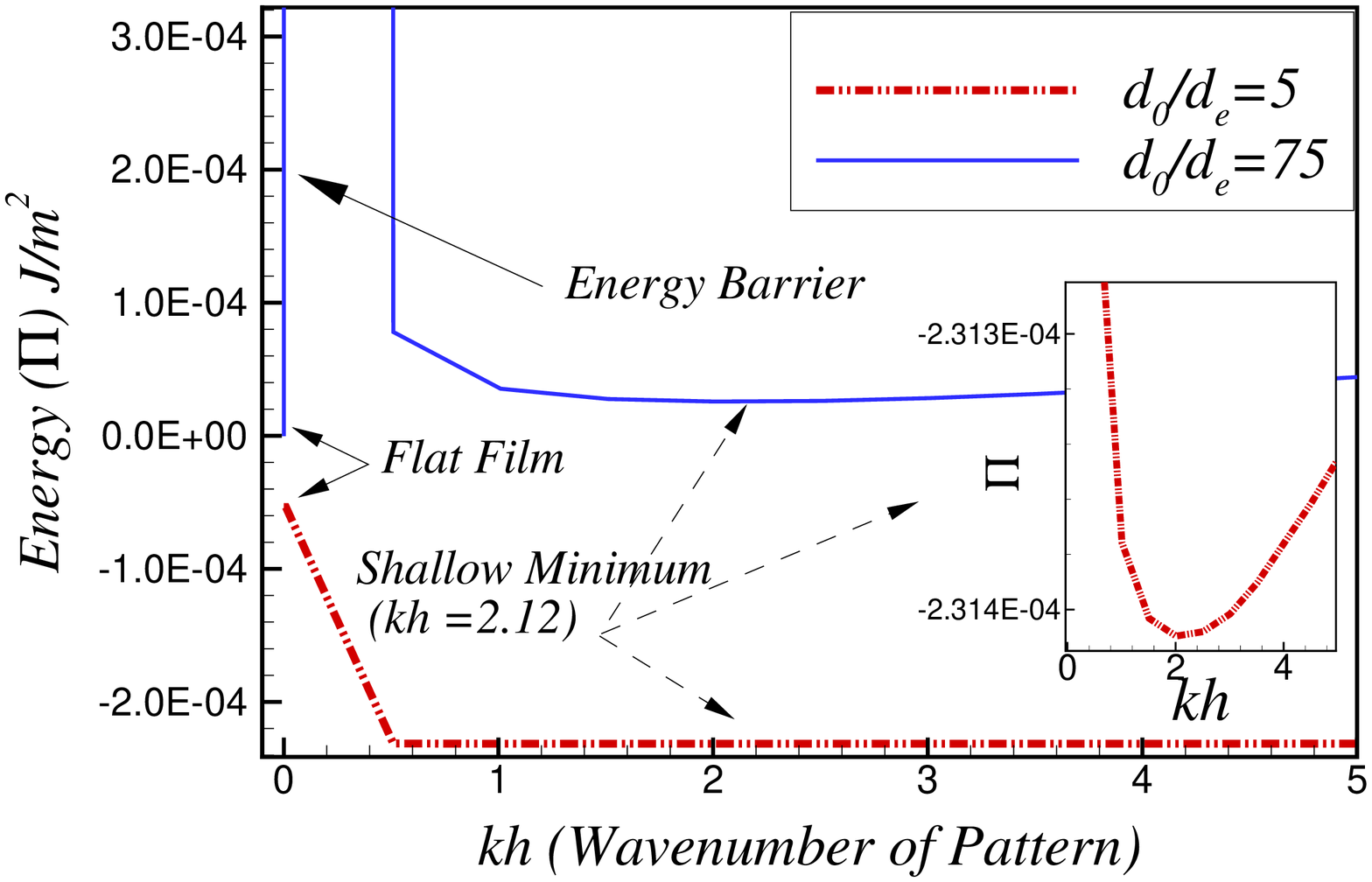}}
\caption{Energy landscape as a function of wavenumber. The plots show
the presence of an energy barrier separating the flat film and
patterned state for large values of $d_0/d_e$. 
 The physical parameters are $\mu = 0.1$MPa, $h = 10$microns, $A =
10^{-20}$J, $\Delta G = 1.0$mJ/m$^2$. }
\label{barrier}
\end{figure}

We explain the physical basis of the adhesion-debonding hysteresis
(debonding distance $\gg d_c$) by the following simple analysis which
is also in conformity with the detailed simulations presented here.
For a single Fourier mode, $u_2(x) = a_k \cos{kx}$, the total energy
(per unit length of the film) is shown in \fig{barrier} for two
different values of the gap-thickness $d_0$ above and below the
critical distance $d_c$.  As predicted by the linear stability, the
patterned configuration with $hk \approx 2.12$ ($\lambda \sim 3h$) has
the lowest energy for $d_0 < d_c$ rather than flat film. However, for
$d_0 > d_c$, the flat film configuration has the lower energy,
although the patterned state remains a {\em local} minimum, metastable
state. For $d_0 > d_c$, the global (flat film) and the local
(patterned state) minima are separated by a large energy barrier
(\fig{barrier}). {\em It is due to the presence this energy barrier
that the patterned states formed during the approach persists in its
metastable configuration upon withdrawal.} Of course, as shown by
simulations, the pattern is far more complex in that it consists of
many Fourier modes, leading to a multiplicity of metastable states of
varying energies. Thus, during the pull off, the system ``hops''
through a succession of metastable states leading to a strong ``path
dependence'' especially in the presence of heterogeneities, noise etc.

The complete simulation of the pull-off process is achieved by the Fourier representation  $u_2(x_1,0)=\sum_{n=0}^{N-1}a_n \cos(k_n x_1)$, where
$a_n$ is the amplitude of the $n$-th Fourier mode with wavenumber
$k_n(=2\pi n/L)$. The total energy per unit depth of the film
\bea
\!\!\!\Pi(a_1,.\!\!\!\!\!&.&\!\!\!\!\!.,a_n)\!\!=\!\!\frac{\pi\mu L}{2}\!\!\sum_{n=0}^{N-1}na_n^2 k_n S(k_n h) \non\\
&+&\!\!\!\!\!\int_0^L\!\!\!U(d_0\!\!-\!\!\sum_{n=0}^{N-1}\!\!a_ncos(k_nx_1))dx_1. \!\!\!\!\!\!\!\!\!\!\!\!\!\!\
\label{E}
\eea 
The stresses that develop in the film are determined from the Fourier
coefficients as $\sigma_{22}(x_1,\!0)\!=2 \mu \sum_{n=0}^{N-1} a_n k_n
S(k_n h) \cos{k_nx_1}$, where $\!S(\xi)\!\! = \!\! \frac{ \! 1 \! + \cosh{(2 \xi)} + 2
\xi^2}{ \! \sinh{(2 \xi)} - 2\xi} \!$.  $F$ is the average force per unit area 
exerted on the contactor plate to hold the film in equilibrium at a given 
separation.


A conjugate gradient(CG) scheme (which finds the local minimum closest
to the initial configuration) was employed to find the Fourier
coefficients that result in a minimum energy pattern for a given
separation distance $d_0 > d_c$ starting from the contact
proximity. The robustness of the energy minimum thus isolated was
confirmed by small random perturbations of the equilibrium
profile. The separation distance was increased in steps of $s$, taking
the energy minimizing pattern of the previous step as the initial
state in the CG scheme. To uncover the range of possible metastable
pathways, we varied the step size $s$ and, in addition, have
considered cases where the energy minimizing Fourier coefficients (at
$d_0$) are perturbed randomly before being taken as initial choices
for the next step.  The perturbations are introduced by multiplying
each Fourier coefficient with $(1 + r)$ where $r$ is a random number
between $-\epsilon$ and $\epsilon$, where $\epsilon$ is called the
noise amplitude.

\begin{figure}
\centerline{\epsfxsize=8.0truecm \epsfbox{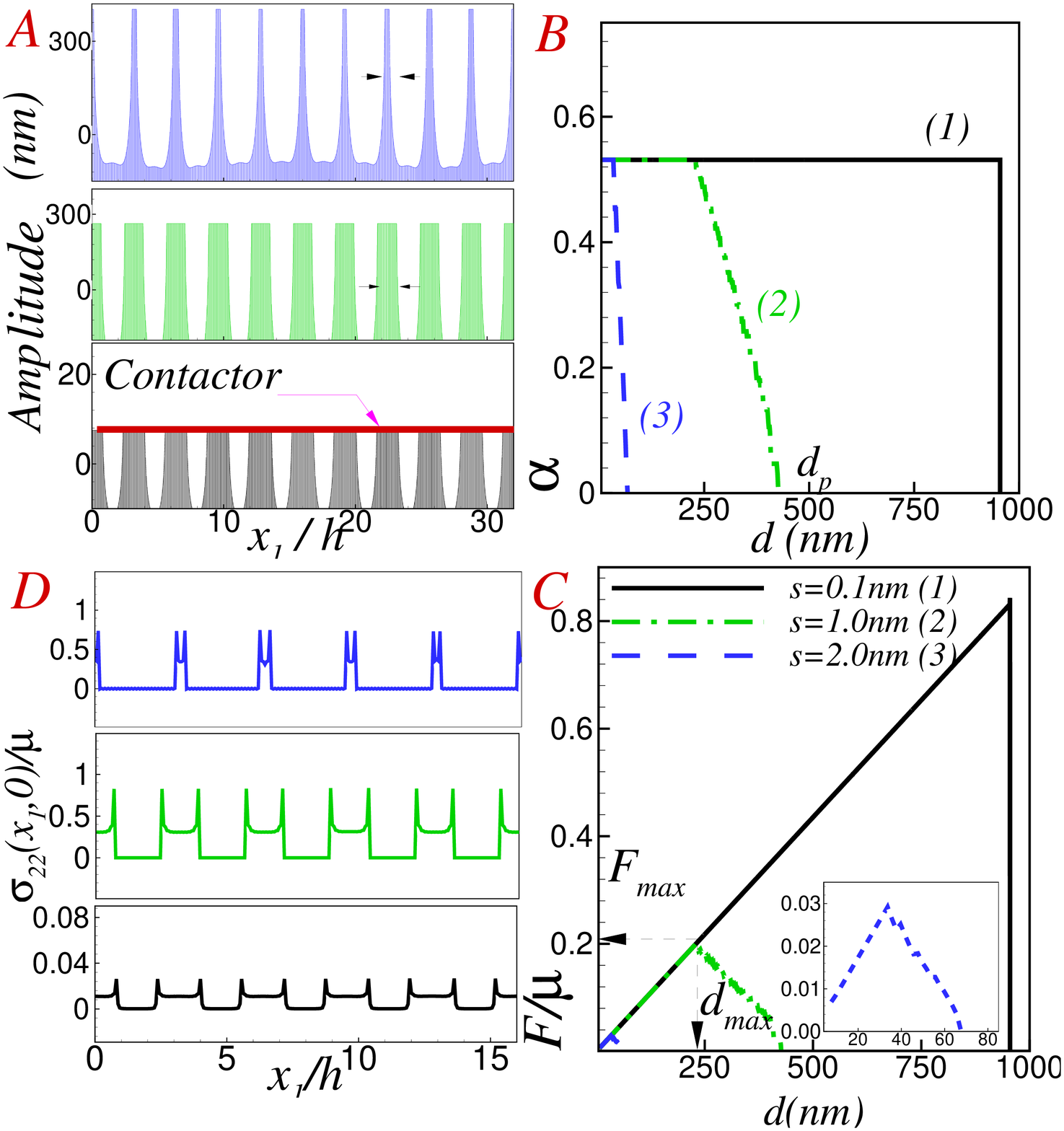}}
\figcap{ (A) Film profiles at various separation distances without external noise and step size of 1nm  ($A=10^{-20}$J, $\mu=0.5$MPa, $h=10$microns) (B) \& (C) Variation, respectively, of fractional contact area $\alpha$ and force $F/\mu$ with separation distances: Curves (1), (2) and (3) correspond to the step size of 0.1nm, 1nm and 2nm, respectively. Curves (1) and (3) correspond to the catastrophic column collapse and continuous peeling modes of failure, and curve (2) is an intermediate. (D) Normal stress distribution along the surface of the film for cases in (A) showing maximum stresses, responsible for peeling, at the column edges.}
\label{Figure2}
\end {figure}

Fig.~\ref{Figure2}A depicts typical changes in the film morphology
during the process of pull off starting from the critical distance
$d_c$ where the instability originates.  This fully nonlinear
simulation (without any imposed noise) shows the columns/cavities 
being laterally separated by $\sim 3h$ at all separation distances until 
the maximum force is reached. Thus, the initial contact $(d<d_c)$ nanocavities predicted by the linear theory persist during pulloff. Figs.~\ref{Figure2}B and \ref{Figure2}C show,
respectively, the variation of contact area ($\alpha =$ area of
contacts/total film surface area) and the force $F$ on the contactor,
with gap thickness for different step sizes, 0.1nm, 1.0nm and
2.nm. For small step sizes ($\lesssim 0.1$nm) the debonding pathway is
such that the configuration is trapped in the energy minimum
corresponding to the initial instability. The contact area remains
constant and the force increases almost linearly until a catastrophic
snap-off of the bridging columns. Remarkably, the maximum force
$F_{max}$ that can be sustained before debonding is about an order of
magnitude smaller than the maximum adhesive force $F_{max}^{flat}$.
  Clearly, debonding does not occur by a uniform
detachment of the contacts, but rather by a different pathway,
requiring much smaller pull off force, made possible by the initial
pattern formation. The formation of bridges and cavities allows very
high concentration of elastic stresses near the edges of the columns.
For small step sizes, the elastic stresses build up to very high
levels comparable to the maximum adhesive force, without any
intermediate small relaxations, since the structure is trapped in the
original deep energy minimum. This engenders a catastrophic adhesive
failure for small step sizes. In contrast, larger step sizes force the
structure to hop through a succession of metastable states with lower
barrier heights releasing energy intermittently leading to a
continuous decrease in the contact area. The stresses at the edges of
the contacts are not large enough to cause catastrophic detachment, but
are sufficient to sustain peeling. For intermediate step sizes (curve
(2) of fig.~\ref{Figure2}B), the initial phase of pulling reproduces
the features of small step size results, followed by the large step
size behavior. The escape from the initial high barrier state occurs
only after after some stretching of columns leading the release of
pent-up elastic energy.  The ascending branch (``elastic branch'') of
the force curve Fig.~3(C) reflects the linear increase of elastic
stresses in the columns without any change in the contact area. The
initiation of peeling limits the maximum force, after which it
declines (``release branch'') with further increase in the separation
distance and a concurrent reduction in the contact area. The release
branch of the force curve is realizable only in displacement
controlled experiments.

A particularly simple model, which shows the essential physics of the
linear decrease of the area, approximates the total energy as
$\frac{3}{2} \mu u^2(\alpha/(1-\alpha)h) + \alpha U(d_0-u)$
where $\alpha$ is the fractional contact area. For a given $d_0$,
the minimum of energy occurs when the fractional contact area is $
\alpha(d_0)\approx 1- (3 \mu/2h|U(d_e)|)^{1/2} \, d_0$. This linear
decrease in the contact area with increased separation shows that
debonding, even by the application of a purely normal force, actually
proceeds by {\em peeling of the contacts}. Peeling from the contact
edges requires much smaller energy penalty (and force) as compared to
homogeneous debonding of  flat contact areas.

\begin{figure}
\centerline{\epsfxsize=8.0truecm \epsfbox{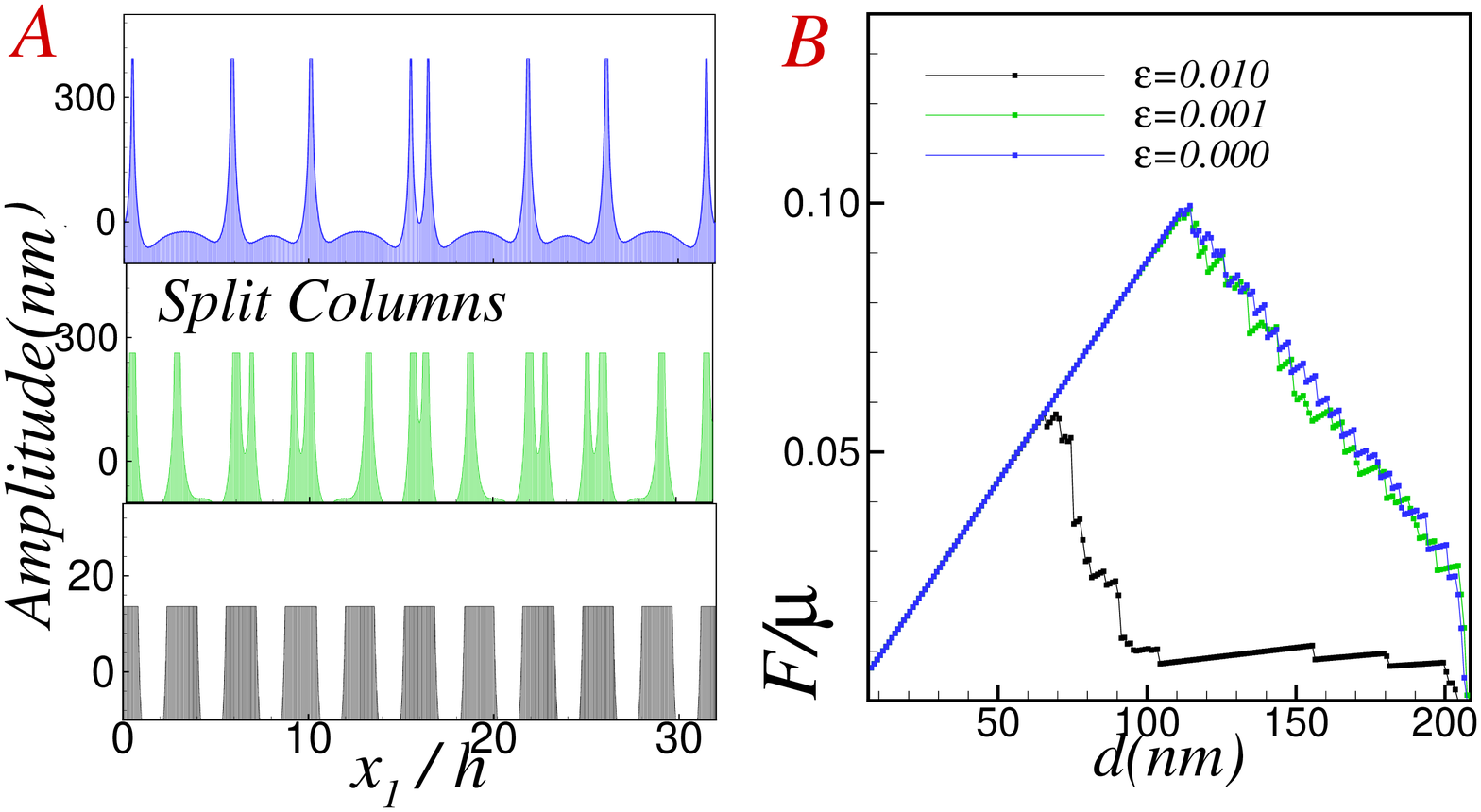}}
\figcap{(A) Noise-induced, column-splitting pathway of adhesive failure for a more strongly adherent film (same as fig.~\ref{Figure2} except, $A=10^{-19}$J and $\epsilon=0.01$) (B) Variation of force $F/\mu$ with separation distances for various levels of noise ($\epsilon=0.0,0.001,0.01$).}
\label{Figure4}
\end {figure}

Although the peeling mode remains the dominant mode of debonding for
small levels of noise, another pathway of debonding in the form of
cavitation within the contact area leading to column splitting also
appears for high noise amplitudes (fig.~\ref{Figure4}A), i.~e.,
starting from initial conditions that are far from the solution branch
being followed in the absence of noise. The column splitting mode is
favored for higher adhesive strength and for more compliant films
(higher value of $\Delta G h/\mu$) where even smaller amounts of noise
can induce this transition. Column splitting, when it occurs, results
in precipitous decrease in the force, usually followed by a regime of more
nearly constant force (fig.~\ref{Figure4}B). Continuous peeling from
the sides of the split columns prevents the build-up of elastic force
in the constant force regime. This helps understand experimental
observations of constant force regime which becomes more prominent on
rough surfaces that allow cavity initiation within the contact zones \cite{3,6}. 

\begin{figure}
\centerline{\epsfxsize=8.0truecm \epsfbox{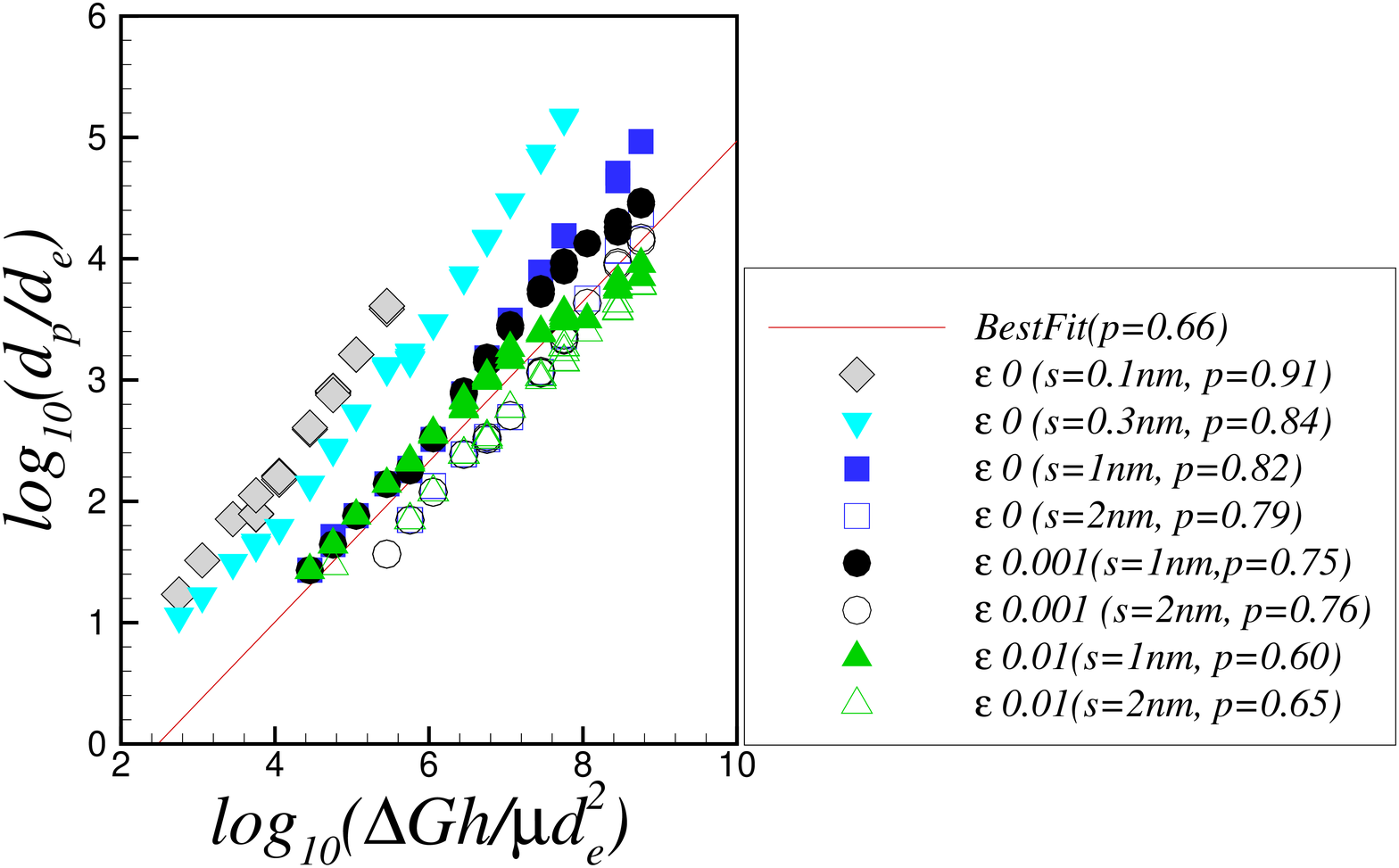}}
\figcap{Variation of snap-off distance with non-dimensional 
stiffness ratio. The slope of each curve is represented by $p$. }
\label{dpcorr}
\end{figure}

The maximum pull-off distance was obtained for a wide range of
parameters $h \sim 0.1-50\mu m, A \sim 10^{-19} - 10^{-21}J$ and $\mu
\sim 0.1 - 10MPa$, step sizes $(s \sim 0.1 - 2.0)$nm, noise
amplitude $(\epsilon\sim 0.001 - 0.01)$ and $\Delta G \sim 1-100$mJ/m$^2$. Interestingly, the dependence of the pulloff distance $d_p$ on $\Delta G$, $\mu$ and $h$ is represented by a master curve of the form (fig.~\ref{dpcorr})
\bea
\frac{d_p}{d_e} \sim \left(\frac{\Delta G h}{\mu d_e^2}\right)^{p}
\label{2}
\eea 
where the nondimensional parameter $(\Delta G/d_e^2)/(\mu/h)$ is
the ratio of the stiffness of the interaction potential and the
elastic stiffness of the film. The exponent $p$ is close to 1 for
noiseless cases with small step size and decreases with increasing
step size as well as increasing level of noise (the minimum exponent
is about 0.6). As argued earlier, increased step size and noise levels
can induce debonding at smaller distance by cascading through higher
energy metastable states.

Further, the maximum force from simulations was found to scale as
\bea
\frac{F_{max}}{\mu} = C \left(\frac{h}{d_e}\right)^\gamma \left(\frac{\Delta G h}{\mu d_e^2}\right)^{\delta}
\label{3}
\eea
where exponents $\gamma,\delta$ are $-0.63$ and $0.24$, respectively
for noiseless case with small step size ($s = 0.1$nm), and the
prefactor $C =38.5$. In this case, $F_{max} \sim (\Delta G)^{0.25}
h^{-0.4} \mu^{0.75}$.  In all other cases of higher step size and noise
considered, $\gamma$ is found to be remarkably constant at $-1\pm0.01$
and the exponent $\delta$ is $0.8\pm0.18$. The prefactor for these
exponents is in the range of $0.01$ to $0.1$. The force in this case
scales as $F_{max} \sim (\Delta G)^{0.8} h^{-0.2} \mu^{0.2}$ showing a
stronger dependence on adhesive energy but a weaker dependence on film
thickness and shear modulus compared to the noiseless and small step
size case. The above considerations (\fig{dpcorr}) also explain the long 
debated contention that the surface energy of soft solids as measured 
from debonding experiments is a non-equilibrium and non-unique property.

The values of the maximum force per unit area required for debonding
are much smaller than predicted for debonding for flat surfaces
($F_{max}^{flat} \sim \Delta G/d_e$). For example with
$A=10^{-20}J,h=10.0\mu m,\mu=0.5MPa$), $F_{max}^{flat} = 80$MPa,
$F_{max} = 0.1$MPa for step size 1nm without noise and $F_{max} =
0.04$MPa with noise. The ratio, $F_{max}/F^{flat}_{max} \sim (\mu
d_e^2/\Delta G h)^n (h/d_e)^m$, $(n>m)$, where $n$ is close to 0.2 and
$m$ close to 0 for cases with noise and relatively large steps. This
shows the discrepancy in the forces between the flat and instability
controlled modes of failure increases with decreasing shear modulus,
increasing adhesive strength and film thickness.

\begin{figure}
\centerline{\epsfxsize=8.0truecm \epsfbox{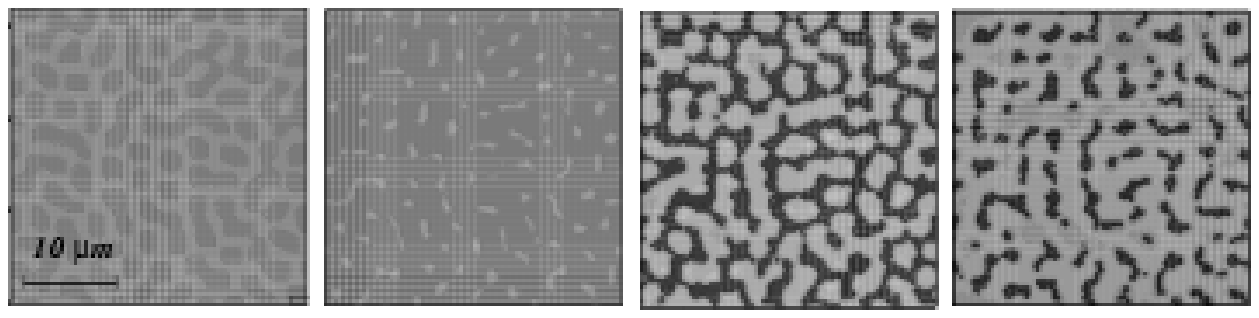}}
\figcap{The first two frames represent the instability pattern during approach
$(d_0 = 5.72$nm and $1.43$nm) and the last two represent pattern
evolution during withdrawal ($d_0 =35.43$nm and 43.43nm) The darker
regions are the contact zones and the lighter shades denote the
cavities formed within the film. During the approach a labyrinth
pattern is transformed into isolated nanocavities which upon pulloff
grow and fragment the contact zones. The wavelength of the pattern
($\sim3h$) also agrees with linear theory and experimental observations \cite{2,3,4}.
}
\label{Figure6}
\end{figure}
Representative 2D simulations (such as shown in fig.~\ref{Figure6})
also confirm the underlying physics and the other results reported here. A
detailed account will be published elsewhere.

This letter resolves important open questions regarding the mechanisms
and pathways of debonding at soft interfaces.  The main results
include, (a) the physical origins of the adhesion-debonding
hysteresis, (b) formation and persistence of regularly arranged
cavities and bridges during debonding, (c) metastable pathways of
debonding such as column collapse, column peeling and column splitting
that require much larger pulloff distances and much smaller debonding
forces as compared to detachment of flat surfaces, (d) complete 
quantitative dependance of pull-off distance and force on adhesion 
energy, shear modulus and film thickness. Formation of
cavities engenders extremely high stresses near the column edges
leading to the peeling of contact zones at much smaller average
stresses than the adhesive strength. This is analogous to defects
(dislocations and cracks) in solids which give rise to observed yield
stress and strength much smaller than ideal values.

Discussions with M.K. Chaudhury and A. Ghatak are gratefully acknowledged.
V.~S.~and A.~S.~acknowledge financial support of DST, India through the Nanoscience program.


\begin{thebibliography}{1}
\bibitem [1]  {1}Y. Y. Lin, C. -Y. Hui and H. D. Conway, J. Poly. Sci. Part B: Poly. Phy., {\bf 38}, 2769 (2000); C. Creton, J. Hooker, and K. R. Shull, Langmuir, {\bf 17}, 4948-4954 (2001).
\bibitem [2] {2} A. Ghatak, M. K. Chaudhury, V. Shenoy and A. Sharma, Phys. Rev. Let., {\bf 85}, 4329 (2000); A. Ghatak and M. K. Chaudhury, Langmuir, {\bf 19}, 2621 (2003).
\bibitem [3] {3} K. R. Shull, C. M. Flanigan and A. J. Crosby, Phys. Rev. Let., {\bf 84}, 3057 (2000); R. E. Webber, K. R. Shull, A. Roos, and C. Creton, Phys. Rev.E.,{\bf 68}(2), 021805 (2003)
\bibitem [4] {4} W. M\"onch,  and S. Herminghaus, Euro. Let., {\bf 53}, 525 (2001).
\bibitem [5] {5} V. Shenoy and A. Sharma, Phy. Rev. Let., {\bf 86}, 119 (2001); J. Mech. and Phy. of Sol., {\bf 50}, 1155 (2002); J. Sarkar, V. Shenoy and A. Sharma, Phy Rev E, {\bf 67}, 031607 (2003)., C. Q. Ru, J. Appl. Phys., {\bf 90}, 6098 (2001).
\bibitem [6] {6} A. Chiche, P. Pareige and C. Creton, C. R. Acad, Sci IV -Phys {\bf 1} (9), 1197 (2000).

\end{thebibliography}
\end{document}